# DC-UNet: Rethinking the U-Net Architecture with Dual Channel Efficient CNN for Medical Images Segmentation

Ange Lou[1], Shuyue Guan[2], Murray Loew[2]

Medical Imaging and Image Analysis Laboratory

[1]Department of Electrical and Computer Engineering, [2]Department of Biomedical Engineering

The George Washington University Medical Center

Washington DC, USA

{angelou, frankshuyueguan, loew}@gwu.edu

*Abstract* — **Recently, deep learning has become much more popular in computer vision area. The Convolution Neural Network (CNN) has brought a breakthrough in images segmentation areas, especially, for medical images. In this regard, U-Net is the predominant approach to medical image segmentation task. The U-Net not only performs well in segmenting multimodal medical images generally, but also in some tough cases of them. However, we found that the classical U-Net architecture has limitation in several aspects. Therefore, we applied modifications: 1) designed efficient CNN architecture to replace encoder and decoder, 2) applied residual module to replace skip connection between encoder and decoder to improve based on the-state-of-the-art U-Net model. Following these modifications, we designed a novel architecture--DC-UNet, as a potential successor to the U-Net architecture. We created a new effective CNN architecture and build the DC-UNet based on this CNN. We have evaluated our model on three datasets with tough cases and have obtained a relative improvement in performance of 2.90%, 1.49% and 11.42% respectively compared with classical U-Net. In addition, we used the Tanimoto similarity to replace the Jaccard similarity for gray-to-gray image comparisons.**

*Keywords* — *convolution neural network; MultiResUnet; deep-learning; medical image segmentation; computer aided diagnosis; DC-UNet.*

## I. INTRODUCTION

The goals of medical image analysis are to provide efficient diagnostic and treatment process for the radiologists and clinicians [1]. Medical imaging devices such as X-ray, CT and MRI can provide information of disease, abnormality and anatomic inside of human body nondestructively. Due to large amount of data and noises interference in medical images, it is important to process images and extract effective information [2] from them. Medical image processing has contributed a lot in medical applications; for example, image segmentation, image registration and image-guided surgery are widely used in medical treatment.

The most important technique in medical image processing is image segmentation, which is to minimize the region of interest (ROI) through some automatic and semi-automatic methods [3]. There are many traditional algorithms are designed to segment tissues or body organs. These methods can be classified as: region based, edge based, threshold and feature based clustering [4].

Region based algorithms are to find a group of connected pixels which have similar properties. Segmentation based on edge detection refers to the boundaries where have an abrupt change in the intensity or brightness value of the image. There are various edge detection algorithms like Sobel detector [5], Canny detector [6] and Fuzzy Inference System (FIS) [7]. Generally, before applying the edge detection, we need to pre-process and enhance the images.

Thresholding is a simple and powerful technique for segmenting images having big contrast of objects and background. Thresholding operation converts a multi-level gray image into a binary image through an appropriate threshold T and divides image pixels into several regions and separate objects from background. Otsu's [8] is the most popular thresholding method.

Clustering is the process to group objects based on some similar properties so that each cluster contains similar objects. Examples of clustering methods are K-means [9], Fuzzy C Means (FCM) [10] etc.

These traditional segmentation methods, however, does not suit with challenge tasks. Take CVC-ClinicDB [45] dataset for example, the shape, size and boundary of polyps are totally different, some polyps with vague boundaries cannot be detected by tradition segmentation approach. Recently, deep learning methods have been shown to outperform previous state-of-the-art machine learning techniques for several computer vision tasks [11]. According to the main principle of recent deep learning segmentation methods, they contain two categories: region-based semantic segmentation and FCN-based semantic segmentation [12].

For region-based deep learning method, Region with CNN feature (RCNN) [13] is a well-known model. RCNN utilizes features from CNN detector to address those complicated tasks. Moreover, any CNN structures can be used as a detector in RCNN model, such as AlexNet [14], VGG [15], GoogLeNet [16] and ResNet [17].

The FCN-based [18] is a pixel-wise segmentation method. Compared with region-based methods, FCN-based does not need to extract the region proposals. Fully convolution network extends from the classical CNN and uses a decoder-like part to generate the segmentation mask. After FCN, large numbers of encoder-decoder models have been applied to segmentation, such as SegNet [19] and U-Net [20]

The U-Net and U-Net-like models have been successfully used to segment various biomedical images, such as liver [21], skin lesion [22] and vessels [23].

In this paper, we develop a novel model called Dual Channel U-Net (DC-UNet), it is an enhanced version of U-Net, which is the most popular and successful deep learning model for biomedical image segmentation so far. And we believe DC-UNet will significantly contribute to the medical image segmentation. We tested the DC-UNet by using a variety of medical images. Our results show that DC-UNet surpasses the classical U-Net model in all the cases, even if DC-UNet has slightly fewer parameters.

II. METHODS

A. *Overview of U-Net Architecture*

Similar to FCN and SegNet, U-Net uses convolutional layers to perform the semantic segmentation. The network is symmetric and can be divided to two parts: encoder and decoder. The encoder follows the typical architecture of a convolutional network, which is used to extract spatial features from the images. A convolution block involves a sequence of two $3 \times 3$ convolution operations, followed by a max-pooling operation with a pooling size of $2 \times 2$ and stride of 2. This block is repeated four times; and, after each down-sampling, the number of filters in the convolution is doubled. Finally, a progression of two $3 \times 3$ convolution operations connect the encoder to the decoder.

The decoder is used to construct the segmentation map from the encoder features. The decoder uses a $2 \times 2$ transposed convolution operation [24] to up-sample the feature map and reduce the feature channels to half at the same time. Then a sequence of two $3 \times 3$ convolution operations is performed again. As the encoder, this succession of up-sampling and two convolution operations is repeated four times, reducing the number of filters to half at each stage. Finally, a $1 \times 1$ convolution operation is performed to generate the final segmentation map. All convolutional layers in the U-Net use the ReLU (Rectified Linear Unit) as activation function [25], except the final layer uses a $1 \times 1$ convolutional layer, and Sigmoid activation function.

Moreover, U-Net architecture introduces a skip connection to transfer the output from encoder to decoder. These feature maps are concatenated with the output of up-sampling operation; and the concatenated feature map is propagated to the successive layers. The skip connections allow the network to retrieve the spatial features lost by pooling operations. The U-Net architecture is illustrated in **Fig. 1.**

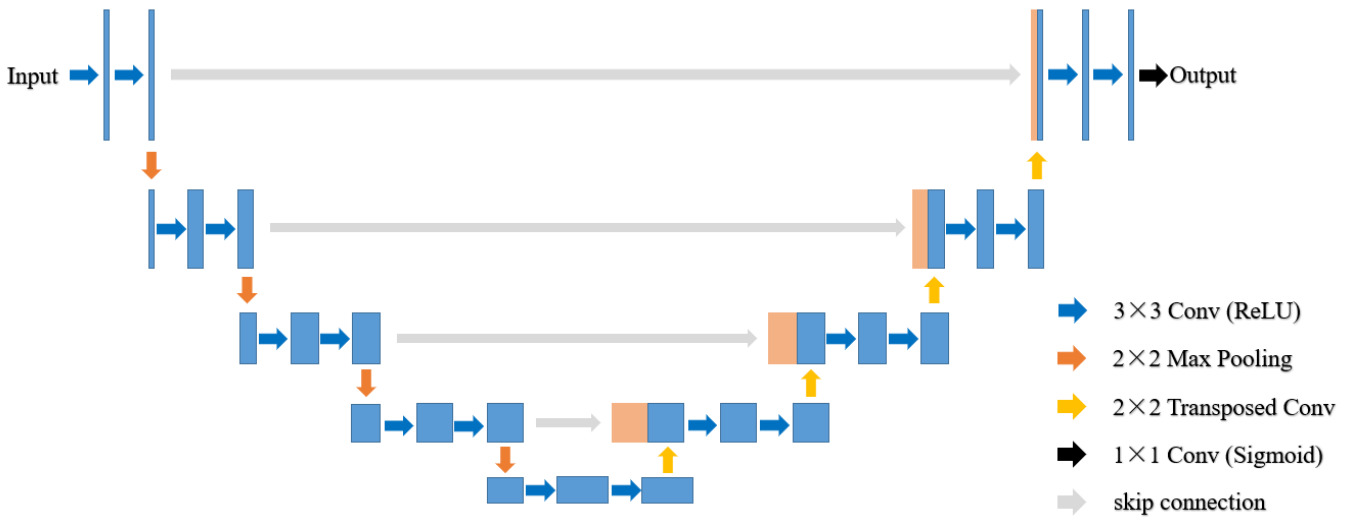

Fig. 1. Architecture of U-Net

## B. Overview of MultiResUNet Architecture

In the medical images, the objects in which we are interested sometimes have different scales. For example, the scale of skin lesions can greatly vary in dermoscopy images. We can find these problems in different medical image segmentation tasks.

Therefore, for better segmentation results, a network needs have the ability to analyze objects at different scales. Previous researcher applied a sequence of Gabor filters with varying scales to acknowledge the variation of scale in the images [26]. Based on this idea, Szegedy [27] introduced a revolutionary architecture — inception blocks. The inception blocks utilize convolutional layers of varying kernel size in parallel to extract features with different scales from images. The inception block is illustrated in **Fig. 2**. In the naïve version, the inception block simply combined $1 \times 1$, $3 \times 3$, $5 \times 5$ convolutional layers and $3 \times 3$ max pooling layers in parallel. Then, it concatenated different scales features and sent them to next layer. One big problem in this naïve version, however, is the number of dimensions will cause a computational blow up. Also, the merging of output of the pooling layer with outputs of the convolutional layers will increase the number of outputs from block to block. The dimension reduction version as shown in **Fig.2 (b) solves the problems**. A $1 \times 1$ convolutional layer [28] are used to reduce dimensions before computing the $3 \times 3$ and $5 \times 5$ convolutions.

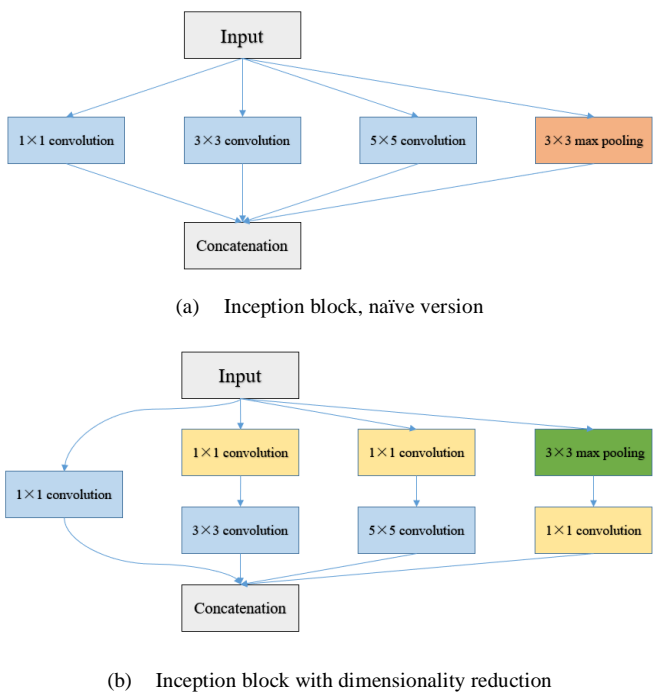

(a) Inception block, naïve version

(b) Inception block with dimensionality reduction

Fig. 2. Inception block

Although using $1 \times 1$ convolutional layer to reduce dimensions, convolution with larger spatial filters (e.g. $5 \times 5$ or $7 \times 7$) is also time-consuming. For example, a $7 \times 7$ convolution is 49 / 9 = 5.44 times more computationally expensive than a $3 \times 3$ convolution with same filter number. Using three $3 \times 3$ convolution layers can obtain a same receptive field output with a $7 \times 7$ convolution [29] but this sequence $3 \times 3$ convolutions are only 27 / 9 = 3 times than a $3 \times 3$ convolution with same filter number. It will be the same with $5 \times 5$ convolution. Based on the replacement, the inception block can also be simplified as in **Fig. 3**.

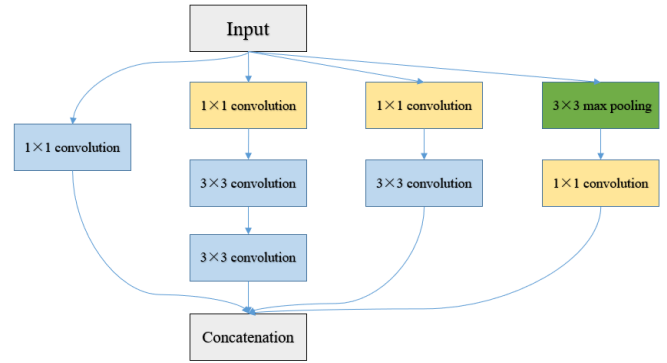

Fig. 3. Inception block where each $5 \times 5$ convolution is replaced by two $3 \times 3$ convolutions.

In the U-Net architecture, after each pooling and transposed convolutional layer, a sequence of two $3 \times 3$ convolutional layers, which can be consider as a $5 \times 5$ convolution, is applied. Like the inception block, to incorporate $3 \times 3$ and $7 \times 7$ convolution operations in parallel to the $5 \times 5$ convolution operation makes the U-Net have multi-resolution analysis ability (**Fig. 4 (a)**).

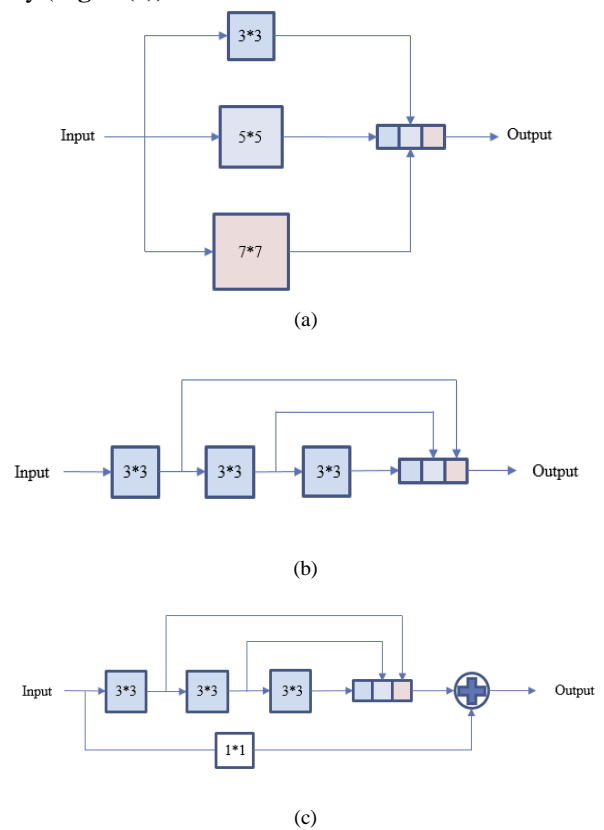

(a)

(b)

(c)

Fig. 4. MultiRes block. (a) A simple inception-like block by using $3 \times 3$, $5 \times 5$ and $7 \times 7$ convolutional filters in parallel and concatenating the generated feature maps. (b) Using a succession of $3 \times 3$ filters to simplify inception-like block. (c) Add a residual connection to build MultiRes block.

To apply this inception-like block makes U-Net architecture have the ability to concatenate features learnt from the image at different scales. In order to reduce computation and memory requirement, we accepted ideas from Szegedy. They utilized a succession of smaller $3 \times 3$ convolutional layers to replace the bigger $5 \times 5$ and $7 \times 7$ convolutional layers, as shown in **Fig. 4 (b)**. Moreover, they also add the $1 \times 1$ convolutional layer called residual connection [30], which can provide some additional spatial features. And this structure is called MultiRes block [31], as shown in **Fig.4 (c)**.

Besides MultiRes block, they also made some modification in skip connection called Res-Path between encoder and decoder. Dataflow pass through a chain of $3 \times 3$ convolutional layers with residual connections, and then concatenate the decoder feature. The Res-Path is illustrated in **Fig. 5**. From the Res-path in stage NO. 1 to NO. 4 in **Fig. 6**, the numbers of $3 \times 3$ convolutional layers with residual connections are {4, 3, 2, 1}.

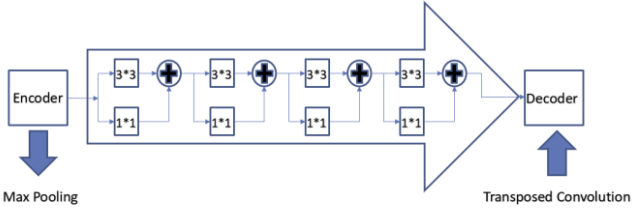

Fig. 5. Res-Path

The MultiRes block and Res-Path form MultiResUNet model, as shown in **Fig. 6**. For the MultiRes block, the filters number of each layer is $W$, which can be computed as:
$$W = \alpha \times U. \qquad (1)$$
$U$ is number of filters in the corresponding CNN block and $\alpha$ is a scalar coefficient.

In the MultiResUNet, the number of filters is equal to {64, 128, 256, 512, 1024}. And we set $\alpha = 1.67$ as a constant. Thus, the numbers of filters for three $3 \times 3$ convolutional layers in the MultiRes are $\frac{W}{6}$, $\frac{W}{3}$ and $\frac{W}{2}$. Moreover, the numbers of filters in the Res-path are {64, 128, 256, 512}. And the numbers of each layers' filters are shown in **Table 1**.

All convolutional layers in the MultiResUNet are activated by the ReLU function and use batch normalization to avoid overfitting. And the final output layer is activated by Sigmoid function.

## C. DC-UNet

### a. Motivation and high-level considerations

The U-Net has been a remarkable and the most popular architecture in medical image segmentation and the MultiResUNet can provide a much better output than the U-Net, because it can provide different scales features. For some extremely challenging medical image cases, however, the MultiResUNet cannot perform well, such as fuzzy objects and interference of backgrounds (part of medical equipment). The goal of MultiRes block is to provide different-scale features to help separate object from the whole image. Hence, we modified the MultiRes block to provide more effective features. This idea led us to build a new block for improvement.

### b. Modification

Our previous work (using MultiResUNet to segment breast region from infrared images) have shown that some small breast IR images do not have clear breast boundary and some segmentation results will be influenced by other interferences, such as patients' belly and parts of medical equipment. Those factors influenced the segmentation results from MultiResUNet. We solved this problem by designing a more effective CNN architecture to extract more spatial features.

To compare the segmentation results from the classical U-Net and MultiResUNet, we can easily find that different-scale features greatly help segmentation. Thus, we assumed that those most challenging tasks would be solved if we can provide more different-scale (more effective) features.

Based on this assumption, we noticed there was a simple residual connection in the MultiRes block. As the author mentioned, the residual connection here only provides a few additional spatial features, which may be not enough to some most challenging tasks. Different-scale feature has already shown the potential in the medical image segmentation. Thus, to overcome the problem of insufficient spatial features, we took a sequence of three $3 \times 3$ convolutional layers to replace the residual connection in MultiRes block. We called this Dual-Channel block, as shown in **Fig. 7**.

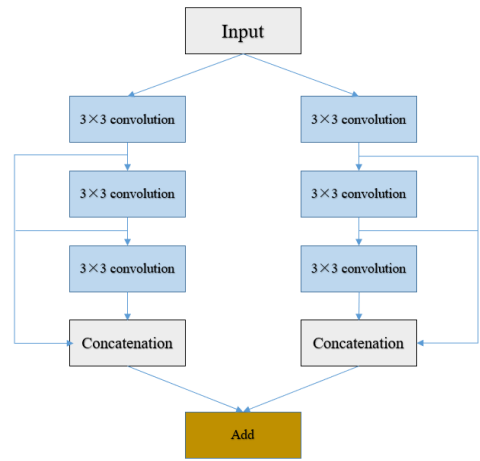

Fig. 7. Dual-Channel block

| MultiResUNet |  |  |  |  |  |
|---|---|---|---|---|---|
| Block | Layer | #Filters | Path | Layer | #Filters |
| MultiRes Block 1<br>MultiRes Block 9 | Conv2D(3,3) | 17 | Res Path 1 | Conv2D(3,3) | 64 |
|  | Conv2D(3,3) | 35 |  | Conv2D(1,1) | 64 |
|  | Conv2D(3,3) | 53 |  | Conv2D(3,3) | 64 |
|  | Conv2D(1,1) | 105 |  | Conv2D(1,1) | 64 |
| MultiRes Block 2<br>MultiRes Block 8 | Conv2D(3,3) | 35 |  | Conv2D(3,3) | 64 |
|  | Conv2D(3,3) | 71 |  | Conv2D(1,1) | 64 |
|  | Conv2D(3,3) | 106 |  | Conv2D(3,3) | 64 |
|  | Conv2D(1,1) | 212 |  | Conv2D(1,1) | 64 |
| MultiRes Block 3<br>MultiRes Block 7 | Conv2D(3,3) | 71 | Res Path 2 | Conv2D(3,3) | 128 |
|  | Conv2D(3,3) | 142 |  | Conv2D(1,1) | 128 |
|  | Conv2D(3,3) | 213 |  | Conv2D(3,3) | 128 |
|  | Conv2D(1,1) | 426 |  | Conv2D(1,1) | 128 |
| MultiRes Block 4<br>MultiRes Block 6 | Conv2D(3,3) | 142 |  | Conv2D(3,3) | 128 |
|  | Conv2D(3,3) | 284 |  | Conv2D(1,1) | 128 |
|  | Conv2D(3,3) | 427 | Res Path 3 | Conv2D(3,3) | 256 |
|  | Conv2D(1,1) | 853 |  | Conv2D(1,1) | 256 |
| MultiRes Block 5 | Conv2D(3,3) | 285 |  | Conv2D(3,3) | 256 |
|  | Conv2D(3,3) | 569 |  | Conv2D(1,1) | 256 |
|  | Conv2D(3,3) | 855 | Res Path 4 | Conv2D(3,3) | 512 |
|  | Conv2D(1,1) | 1709 |  | Conv2D(1,1) | 512 |

Table 1. Details of MultiResUNet

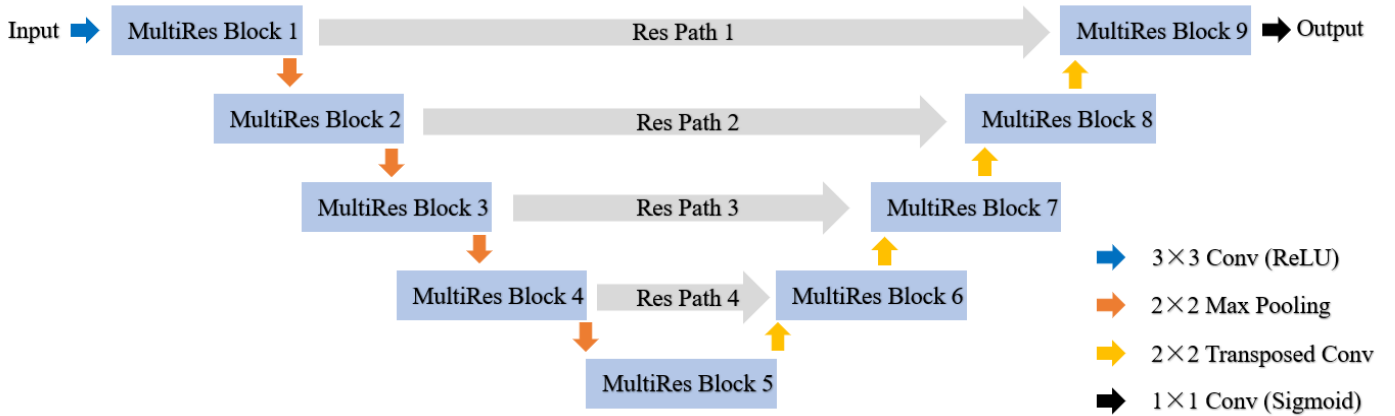

Fig. 6. Architecture of MultiResUNet

We applied the same connection (Res-Path) between encoder and decoder like the MultiResUNet. Then, we utilized the Res-Path and Dual-Channel block to build a new U-Net architecture — DC-UNet whose architecture is illustrated in **Fig. 8**.

Each channel in Dual-Channel block has half filter numbers of MultiRes block :{32, 64, 128, 256, 512}. $W$ is each layers' filter number. And $W$ meets the equation (1) as mentioned before. We applied the same $U$ and $\alpha$ value, and the filter number of three $3 \times 3$ are also divided into $\frac{W}{6}$, $\frac{W}{3}$ and $\frac{W}{2}$. Moreover, the number of filters in Res-path are {32, 64, 128, 256}. And the number of each layers' filter is shown in **Table 2**.

All convolutional layers in the DC-UNet are activated by the ReLU function and use batch normalization to avoid overfitting. And the final output layer is activated by Sigmoid function.

| DC-UNet |||||||||
|---|---|---|---|---|---|---|---|---|
| Block | Layer (left) | #Filters | Layer (right) | #Filters | Path | Layer | #Filters ||
| DC Block 1<br>DC Block 9 | Conv2D(3,3) | 8 | Conv2D(3,3) | 8 | Res Path 1 | Conv2D(3,3) | 32 ||
| | Conv2D(3,3) | 17 | Conv2D(3,3) | 17 | | Conv2D(1,1) | 32 ||
| | Conv2D(3,3) | 26 | Conv2D(3,3) | 26 | | Conv2D(3,3) | 32 ||
| | | | | | | Conv2D(1,1) | 32 ||
| DC Block 2<br>DC Block 8 | Conv2D(3,3) | 17 | Conv2D(3,3) | 17 | | Conv2D(3,3) | 32 ||
| | Conv2D(3,3) | 35 | Conv2D(3,3) | 35 | | Conv2D(1,1) | 32 ||
| | Conv2D(3,3) | 53 | Conv2D(3,3) | 53 | | Conv2D(3,3) | 32 ||
| | | | | | | Conv2D(1,1) | 32 ||
| DC Block 3<br>DC Block 7 | Conv2D(3,3) | 35 | Conv2D(3,3) | 35 | Res Path 2 | Conv2D(3,3) | 64 ||
| | Conv2D(3,3) | 71 | Conv2D(3,3) | 71 | | Conv2D(1,1) | 64 ||
| | Conv2D(3,3) | 106 | Conv2D(3,3) | 106 | | Conv2D(3,3) | 64 ||
| | | | | | | Conv2D(1,1) | 64 ||
| DC Block 4<br>DC Block 6 | Conv2D(3,3) | 71 | Conv2D(3,3) | 71 | | Conv2D(3,3) | 64 ||
| | Conv2D(3,3) | 142 | Conv2D(3,3) | 142 | | Conv2D(1,1) | 64 ||
| | Conv2D(3,3) | 213 | Conv2D(3,3) | 213 | Res Path 3 | Conv2D(3,3) | 128 ||
| | | | | | | Conv2D(1,1) | 128 ||
| DC Block 5 | Conv2D(3,3) | 142 | Conv2D(3,3) | 142 | | Conv2D(3,3) | 128 ||
| | Conv2D(3,3) | 284 | Conv2D(3,3) | 284 | | Conv2D(1,1) | 128 ||
| | Conv2D(3,3) | 427 | Conv2D(3,3) | 427 | Res Path 4 | Conv2D(3,3) | 256 ||
| | | | | | | Conv2D(1,1) | 256 ||

Table 2. Details of DC-UNet

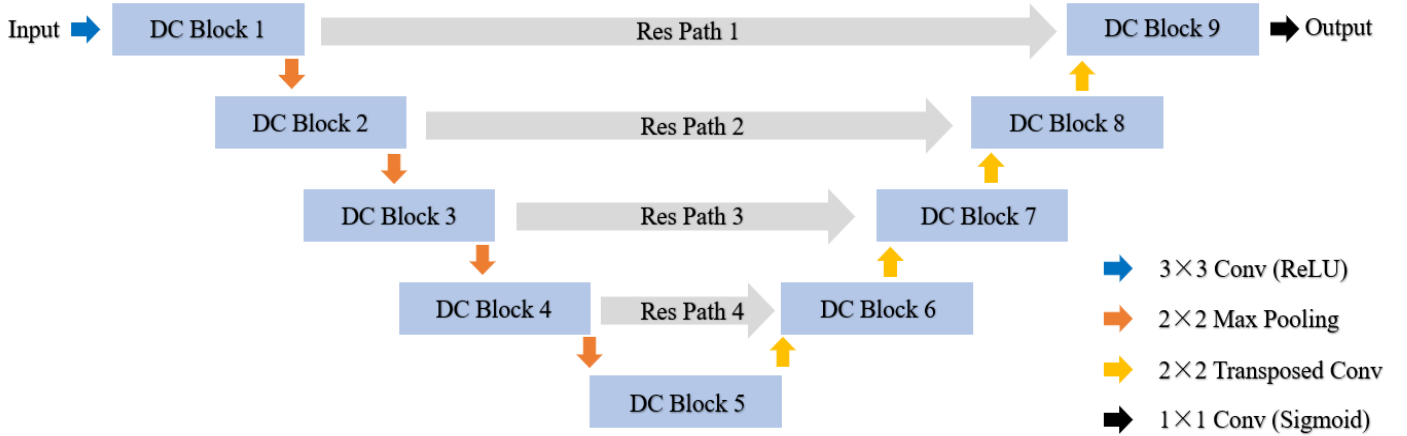

Fig. 8. Architecture of DC-UNet

III. EXPERIMENTS

In the experiments, the network models were built by using Keras [33] with Tensorflow backend [34] in Python 3 [32]. The experiments were conducted in a desktop computer with Intel core i7-9700K processor (3.6 GHz) CPU, 16.0 GB RAM, and NVIDIA GeForce RTX 2070 GPU.

*A. Baseline model*

In these experiments, we chose the U-Net as baseline model and compare its performance with MultiResUNet and DC-UNet. In order to show the advantage in parameters, we implemented the classical U-Net with five stages encoder and decoder, and the filter numbers are {64, 128, 256, 512, 1024}. For the MultiResUNet and DC-UNet, we also set five stages encoder and decoder, and each layer's filter number can be found in **Table 1 and Table 2**.

## B. Pre-processing

The goal of our experiments is to show the performance of the DC-UNet and compare with the classical U-Net and MultiResUNet. The pre-processing we applied for thermography database are converting 16-bit images to 8-bit and resize the image to $256 \times 128$. Due to the limitation of GPU memory, the pre-processing for other databases is to resize the weight and height of images no larger than 256.

## C. Training

The goal of semantic segmentation is to predict whether a pixel belongs to the object. Therefore, this problem can be considered as a pixel-wise binary classification problem. Hence, we chose the binary cross-entropy as loss function and minimized it.

For the input image $X$, the prediction of model is $\hat{y}$ and the ground truth is $y$. Thus, the binary cross-entropy is defined as:

$$Cross\ Entropy(y, \hat{y}) = \sum_{x \in X} -(y \log(\hat{y}) + (1-y) \log(1-\hat{y})) \quad (2)$$

For a batch containing n images, the loss function $J$ is defined as:

$$J = \frac{1}{n} \sum_{i=1}^{n} Cross\ Entropy(y, \hat{y}) \quad (3)$$

We trained those models using the Adam optimizer [35] with the parameter $\beta_1 = 0.9$ and $\beta_2 = 0.999$. Epochs are varied by datasets.

## D. Measurement metric

To evaluate the performance of segmentation, we need a method to compare the segmented region with ground truth region. Since the final layer is activated by a Sigmoid function, it produces output in the range [0, 1]. Therefore, we cannot compare output with ground truth directly, because the ground truth are binary images. Usually, image thresholding from grayscale to binary (binarization) [36] will lose many information. Took our previous infrared breast region segmentation study for example, after converting 16-bit images to 8-bit (pixel value range in [0, 255]), there are lots of usable comparison methods between two images:

- Binary vs Binary: Jaccard Similarity (JS) [37]
- Gray vs Gray: Mean Absolute Error (MAE) [38]; Tanimoto Similarity [39] (Extended Jaccard Similarity); Structural similarity (SSIM) [40]

In our previous studies, we used the JS. JS is for binary to binary comparison; to consider two binary images as two set $A$ and $B$, their JS value is:

$$JS(A, B) = \frac{|A \cap B|}{|A \cup B|} \quad (4)$$

The used conversion method is Otsu [41] based algorithm which is designed by us. The workflow is shown in the **Fig. 9**.

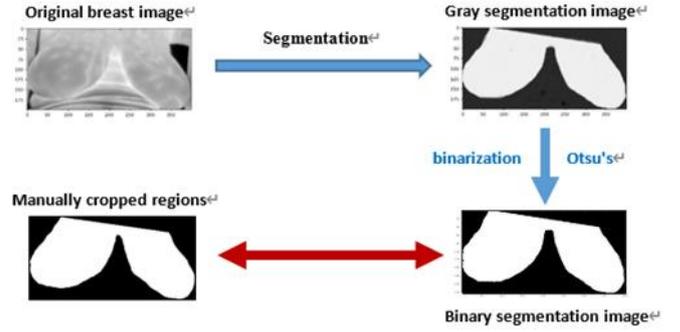

Fig. 9. Workflow of Jaccard Similarity

Otsu's is an automatic image thresholding method to binarize images, however, its outcomes may vary greatly when the breast boundaries of segmented images are not very clear. Hence, we turn to consider some gray to gray comparisons.

A simple and widely used gray to gray comparison based on pixel by pixel comparison is the Mean Absolute Error (MAE). In general, for two images (same size) $A$ and $B$, their MAE value is:

$$MAE(A, B) = 1 - \frac{|A-B|}{maxE} \quad (5)$$

The *maxE* is maximum error value, to 8-bit gray-scale images (size: $W \times L$), $maxE = W \times L \times 2^8$.

The Tanimoto similarity, also called extended JS, can be seen as a grayscale version JS. For binary image, JS compares images by union and intersection operations. The union operation could be considered as sum of products. For two set $A$ and $B$:

$$|A \cap B| = \sum a_i b_i \quad (6)$$

Where $a_i \in A, b_i \in B$. This equation holds if $a_i, b_i \in \{0,1\}$, which are binary values. But if $a_i, b_i$ are not binary, we use sum of products (right part) instead of the union operation. Since:

$$|A \cap A| = \sum a_i^2 \quad (7)$$

And,

$$|A \cup B| = |A| + |B| - |A \cap B| = \sum(a_i^2 + b_i^2 - a_i b_i) \quad (8)$$

For gray-gray comparison, according to $JS(A, B)$, the value of Tanimoto similarity is:

$$T(A, B) = \frac{\sum a_i b_i}{\sum(a_i^2 + b_i^2 - a_i b_i)} \quad (9)$$

By definition, Tanimoto similarity is similar to JS but more general than JS and has wider applications. Therefore, it is a good alternative method for segmentation evaluation.

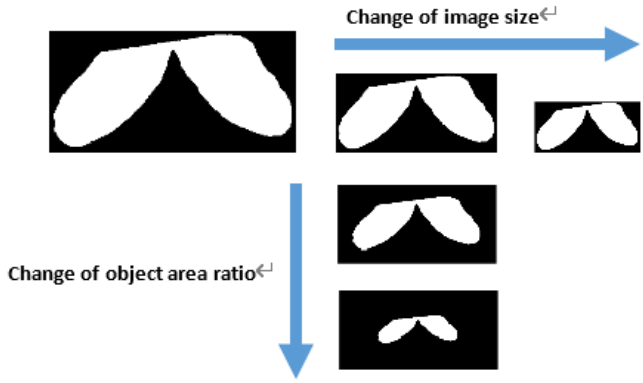

Fig. 10. Size and ratio change of images

Ideal evaluation of images will not be affected by image sizes and object area ratio; as shown in the **Fig. 10**. We tested the JS, MAE, Tanimoto similarity and SSIM on different size images. Each result is the average value of all (15) samples from one patient. For every sample, the value is calculated by comparing ground truth image with C-DCNN segmented image. We changed image size by down-sampling and changed object area ratio by adding blank margin around the object. From **Fig. 11**, results show that SSIM is not stable to image size change and only the Tanimoto similarity keeps stable to changes of object area ratio.

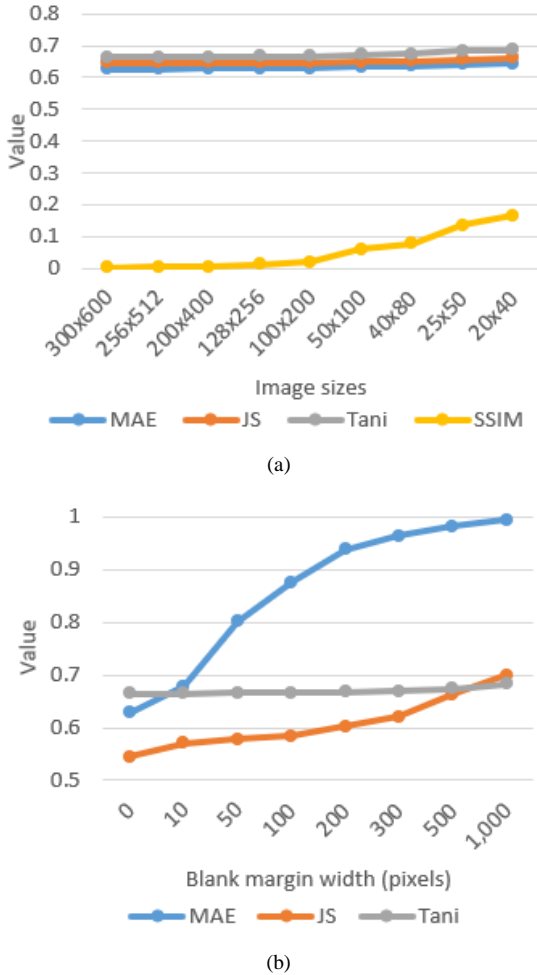

Fig. 11. (a) accuracy value vs image size, (b) accuracy vs ratio

Furthermore, **Fig. 12** shows comparison results by Tanimoto similarity, JS and MAE for the 15 samples in size 200x400. Results indicate that for majority samples (9/15, yellow mark), Tanimoto similarity values are close to JS. Therefore, Tanimoto similarity is a good alternative measure instead of JS for grayscale image comparisons.

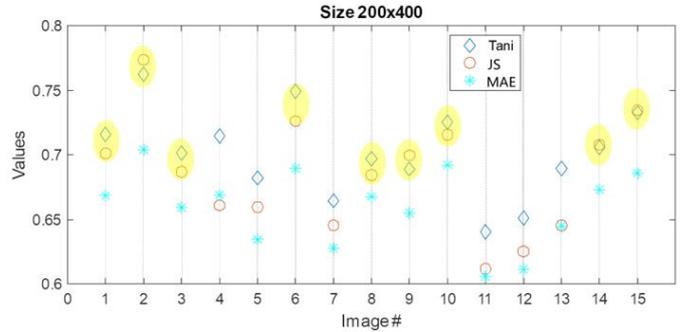

Fig. 12. Comparison of four measurement matric

In conclusion, Jaccard similarity is a proper measure if segmentation output is binary but for grayscale images, Tanimoto similarity is better. In this study, we use Tanimoto similarity to all evaluations for all datasets segmentation because segmented images from neural networks are grayscale, thus, using Tanimoto similarity avoids binarization so that keeps more information of segmented images and time cost is less, Tanimoto similarity is stable on different image size and object ratio, in addition, results are close to JS. Although ground truth images are binary, it is simple to convert them to 8-bit grayscale by multiplying by 255.

*E. Cross-validation*

Cross-Validation is widely used to test model's performance. In the k-Fold cross-validation test, the dataset $D$ is randomly split into $k$ mutually exclusive subsets $D_1, D_2, \ldots, D_k$ of approximately equal size [42]. The model is run k times; for each time, one of the $k$ subsets is chosen as the validation set and all others as training set. We estimated the performance of model via overall results from k times training.

## IV. DATASETS

Compared with traditional computer vision datasets, current medical imaging datasets are more challenging. Expensive medical equipment, complex image acquisition pipelines, diagnose of expert and tedious manual labeling – they all make medical datasets hard to build. Currently, there are some public medical imaging benchmark datasets containing medical images and their ground truth. We have selected two public datasets and our own infrared breast dataset to test the performance of the three U-Net based models. The datasets used in the experiments are briefly described in **Table 3**.

| Modality | Dataset | NO. of images | Original resolution | Input resolution |
|---|---|---|---|---|
| Thermography | Our IR breast | 450 | Variable | $256 \times 128$ |
| Electron microscopy | ISBI-2012 | 30 | $512 \times 512$ | $256 \times 256$ |
| Endoscopy | CVC_ClinicDB | 612 | $384 \times 288$ | $128 \times 96$ |

Table 3. Overview of the datasets.

### A. Infrared breast images

We collected infrared images using the N2 Imager (N2 Imaging System, Irvine, Calif.). Patients diagnosed with breast cancer and healthy volunteers are imaged by the infrared camera for 15 minutes to observe cool down of the breast tissue. This dynamic thermography monitors the temporal behavior of breast thermal patterns, which in our case is the cool down of breast tissue over time. The patients and volunteers keep sitting with both arms raising on two arm supports, with the camera positioned approximately 25 inches away from the breasts (frontal view). Imager starts to capture images immediately after the patient undressed, images were taken every minute along with the breast skin cooling down.

Our breast dataset contains 450 infrared images from 14 patients and 16 healthy volunteers; all images contain background objects and noise. Each participant was imaged for a total time of 15 minutes, capturing one image every minute (15 images per participant). The original resolution of images ranges from $540 \times 260$ to $610 \times 290$; we have resized them to $256 \times 128$ due to limitation of memory.

### B. Electron microscopy (EM)

To show the performance of new model in electron microscopy (EM) images, we choose the dataset of the ISBI-2012 challenge: 2D EM segmentation [43]. This dataset contains 30 images in its training set from a serial section Transmission Electron Microscopy (ssTEM) of the Drosophila first instar larva ventral nerve cord [44]. Due to the testing set does not contain ground truth, we chose totally 30 images in training set as dataset. The resolution of images is $512 \times 512$, we have resized the images to $256 \times 256$ due to the limitation of memory. And for EM segmentation experiment, we took 5-Fold cross-validation.

### C. Endoscopy images

To show the performance of new model in endoscopy images, we chose CVC-ClinicDB [45] as dataset. These images were extracted from the colonoscopy videos. This dataset contains total 612 images with ground truth and their original resolution was $384 \times 288$. We resized the images to $128 \times 96$ for training due to the limitation of memory.

### V. RESULTS

#### A. Models in experiments

To evaluate the performance of DC-UNet, we have designed experiments with three medical datasets and shown the parameter numbers of models in **Table 4**.

| Model | Parameters |
|---|---|
| U-Net (baseline) | 31,031,685 |
| MultiResUNet | 29,061,741 |
| DC-UNet | 10,069,640 |

Table 4. Models used in experiments.

#### B. Results of infrared breast images

The infrared breast dataset contains 450 images and was divided into 30 subsets ($k = 30$) by participant. Every model has been trained 50 epochs at each run. The overall average accuracies of U-Net, MultiResUNet and DC-UNet are 89.80%, 91.47% and 92.71%, respectively, after applying the 30-Fold cross validation for three models. The average accuracies and standard deviations for each participant are shown in **Fig 13**, the DC-UNet performs better than the other models for most test cases. Table 5 shows the specific average accuracy values of each patients.

|  | P1 | P2 | P3 | P4 | P5 | P6 | P7 | P8 | P9 | P10 | P11 | P12 | P13 | P14 | V1 |
|---|---|---|---|---|---|---|---|---|---|---|---|---|---|---|---|
| U-NET | 85.2 | 87.9 | 87.0 | 94.4 | 89.1 | 91.1 | 89.7 | 79.8 | 89.4 | 94.0 | 87.0 | 90.1 | 93.3 | **87.7** | 93.1 |
| MultiResUNet | 88.2 | 92.9 | 90.5 | **96.7** | **91.8** | 91.6 | 89.0 | 80.0 | 90.5 | 95.3 | 84.0 | 95.1 | 93.2 | 87.0 | **95.9** |
| DC-UNet | **90.3** | **94.7** | **91.1** | **96.7** | 90.7 | **92.4** | **90.0** | **83.4** | **91.4** | **95.5** | **92.6** | **96.2** | **95.9** | 87.6 | 95.7 |

|  | V2 | V3 | V4 | V5 | V6 | V7 | V8 | V9 | V10 | V11 | V12 | V13 | V14 | V15 | V16 |
|---|---|---|---|---|---|---|---|---|---|---|---|---|---|---|---|
| U-NET | 92.2 | 90.4 | 92.7 | 88.7 | 85.5 | 92.5 | 84.3 | 95.1 | 82.7 | 94.1 | 94.8 | 89.8 | 91.8 | 93.0 | 87.9 |
| MultiResUNet | 93.0 | 90.8 | 92.6 | 90.3 | 90.1 | 93.9 | 88.2 | 96.0 | 88.4 | 96.5 | **95.7** | 89.6 | 93.2 | 95.2 | 89.0 |
| DC-UNet | **93.8** | **93.6** | **94.1** | **91.4** | **91.1** | **95.4** | **88.4** | **96.9** | **89.3** | **97.1** | 95.5 | **90.3** | **94.1** | **95.5** | **91.0** |

Table 5. Average segmentation accuracy for each sample. Bold values are the maximum for each participant.

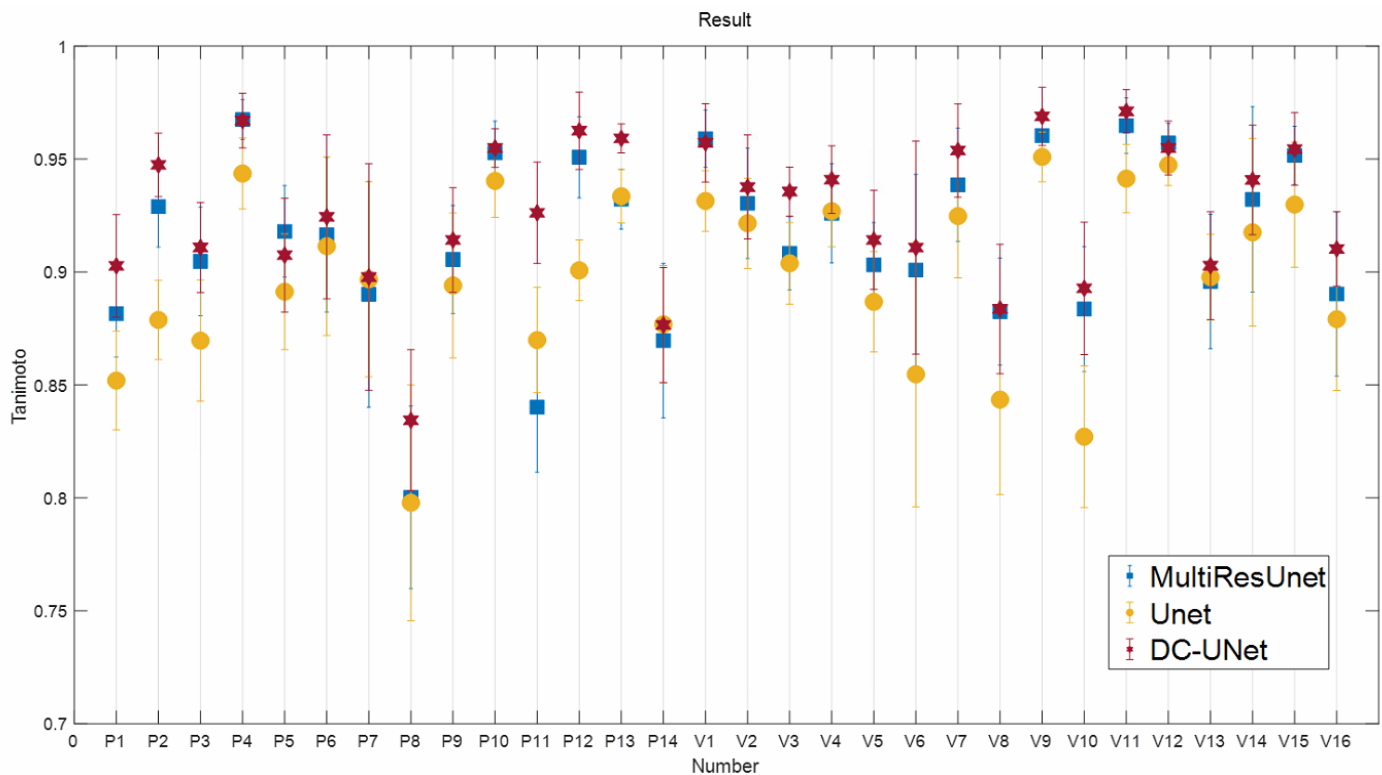

Fig. 13. Breast segmentation results of the models for each subject. The "P" and "V" corresponds to patient and volunteer.

From the results in Table 5, DC-UNet provides more accurate segmentation results both for simple and challenging cases. For example, **Fig. 14** shows a simple case. The segmentation accuracies of U-Net, MultiResUNet and DC-UNet were 92.47%, 93.86% and 95.38%, respectively. DC-UNet gives the best segmentation results compared to the other models because of using our new DC blocks.

results. For the example of patient No.11, its image contains many interferences like medical equipment and other parts of body as shown in **Fig. 15**. The segmentation accuracies of U-Net, MultiResUNet and DC-UNet are 86.47%, 84.01% and 92.62%, respectively. We can find that only DC-UNet can clearly separate breast boundary from belly. Thus, the DC-UNet gives an outstanding result than other models.

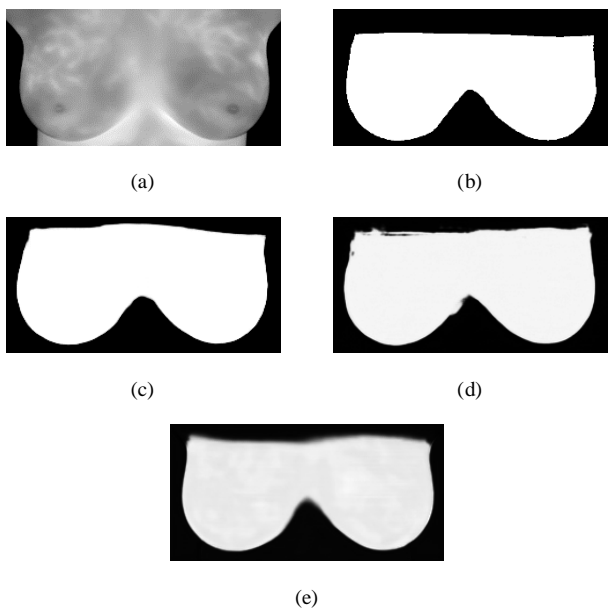

Fig. 14. Segmentation result of volunteer 7. (a) Original image (b) Manual ground-truth (c) U-Net (d) MultiResUNet (e) DC-UNet

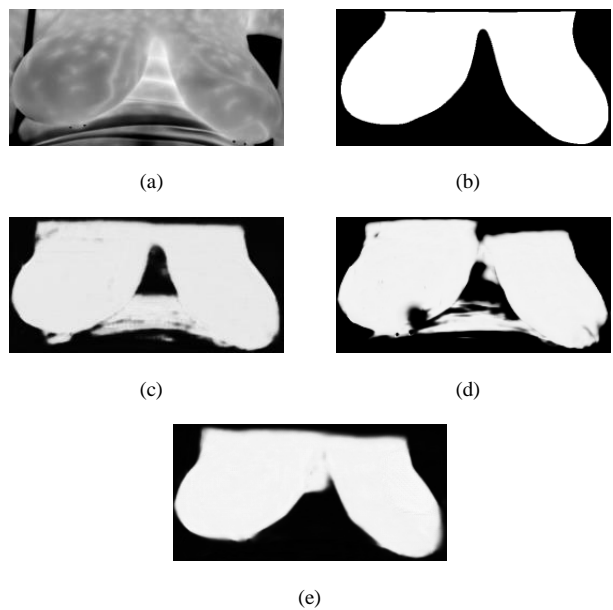

Fig. 15. Segmentation result of patient 13. (a) Original image (b) Manual ground-truth (c) U-Net (d) MultiResUNet (e) DC-UNet

For the challenging cases, DC-UNet also gives inspiring

C. *Results of electron microscopy image*

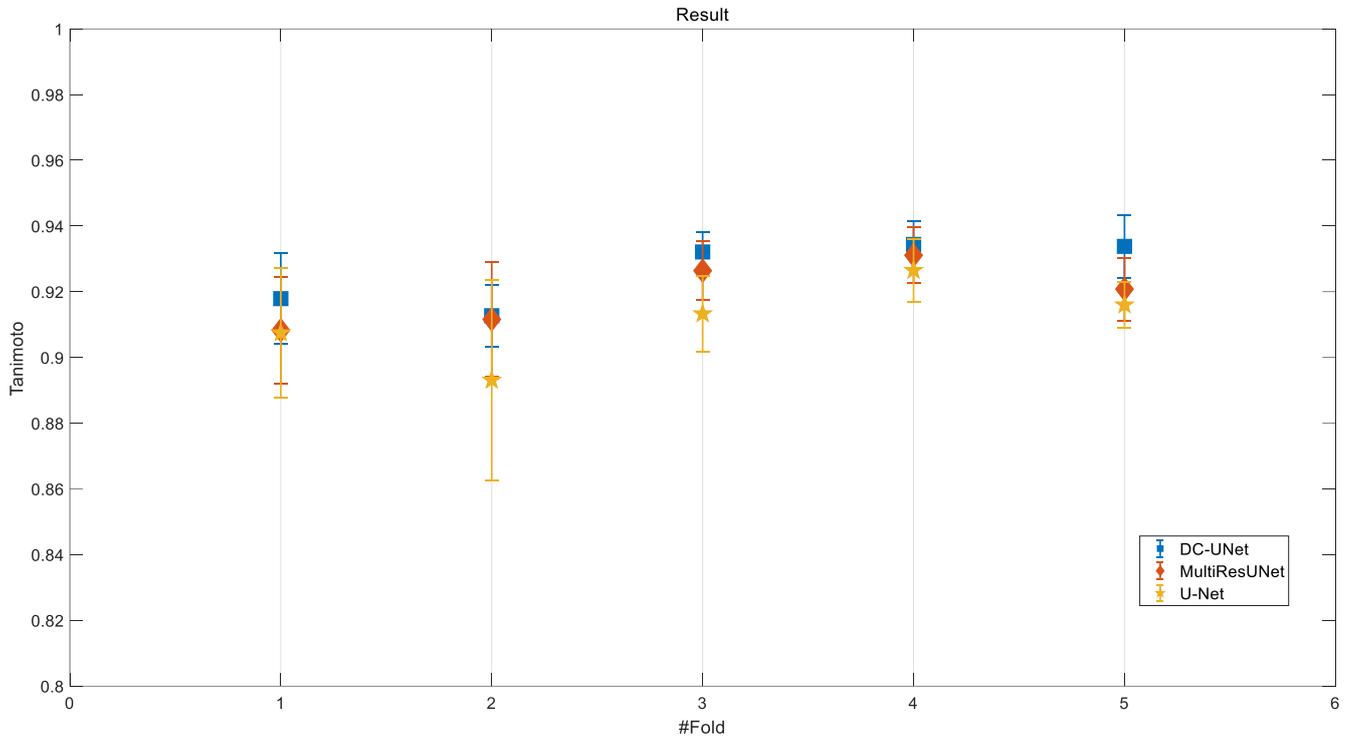

Fig. 17. Segmentation results of the models for each fold

For the electron microscopy (EM) dataset, we have performed 5-fold cross-validation and compared the performance of DC-UNet with MultiResUnet and the baseline U-Net. Every model has been trained 50 epochs for each run and recorded the Tanimoto accuracy. The results of EM dataset were shown in **Table 6**.

| Model | Fold 1 | Fold 2 | Fold 3 | Fold 4 | Fold 5 | Average |
|---|---|---|---|---|---|---|
| U-Net | 90.75 | 89.31 | 91.33 | 92.65 | 91.60 | 91.13 |
| MultiResUNet | 90.83 | 91.16 | 92.64 | 93.11 | 92.08 | 91.96 |
| DC-UNet | **91.79** | **91.27** | **93.21** | **93.44** | **93.38** | **92.62** |

Table 6. The result of EM through a 5-fold cross-validation. Bold values are the maximum for each case.

From the Table 7, we can find that the DC-UNet gives the best results for all cases. Segmentation results of one case are shown in **Fig. 16**. And the average accuracies and standard deviations for each fold are shown in **Fig. 17**.

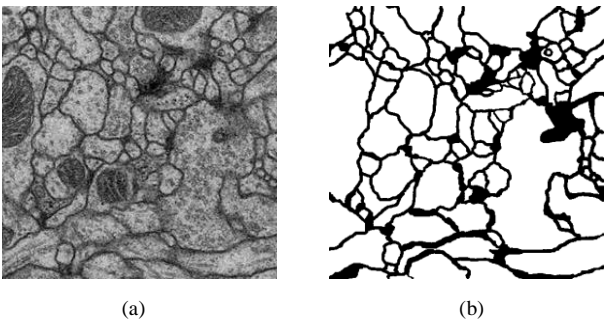

(a)          (b)

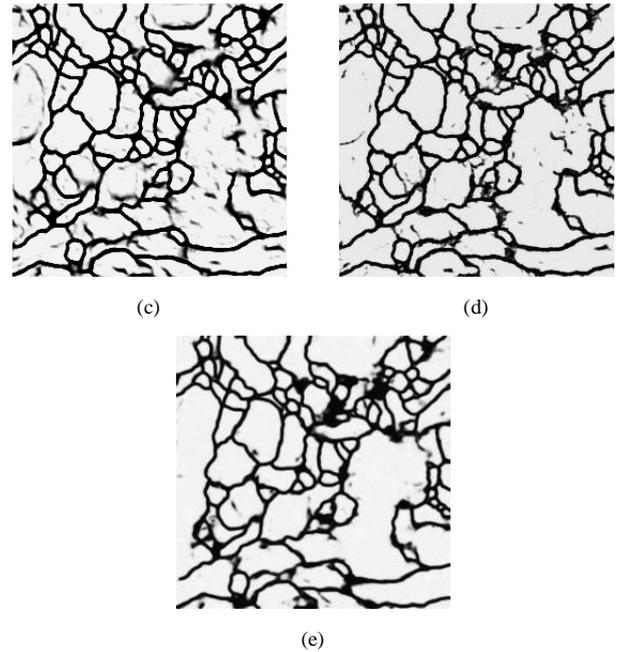

(c)          (d)

(e)

Fig. 16. Segmentation results. (a) Original image (b) Ground truth (c) U-Net (89.8) (d) MultiResUNet (91.4) (e) DC-Net (93.8).

In the Fig. 16, we can find that DC-UNet has a better result than MultiResUNet and the visually segmentation result are very different. The segmentation results show that the DC-UNet can capture some separating lines that U-Net and MultiResUNet missed. Moreover, the DC-UNet can distinguish the boundary from interferences so that the segmentation result of DC-UNet is much clearer than other two models.

D.  *Results of endoscopy image*

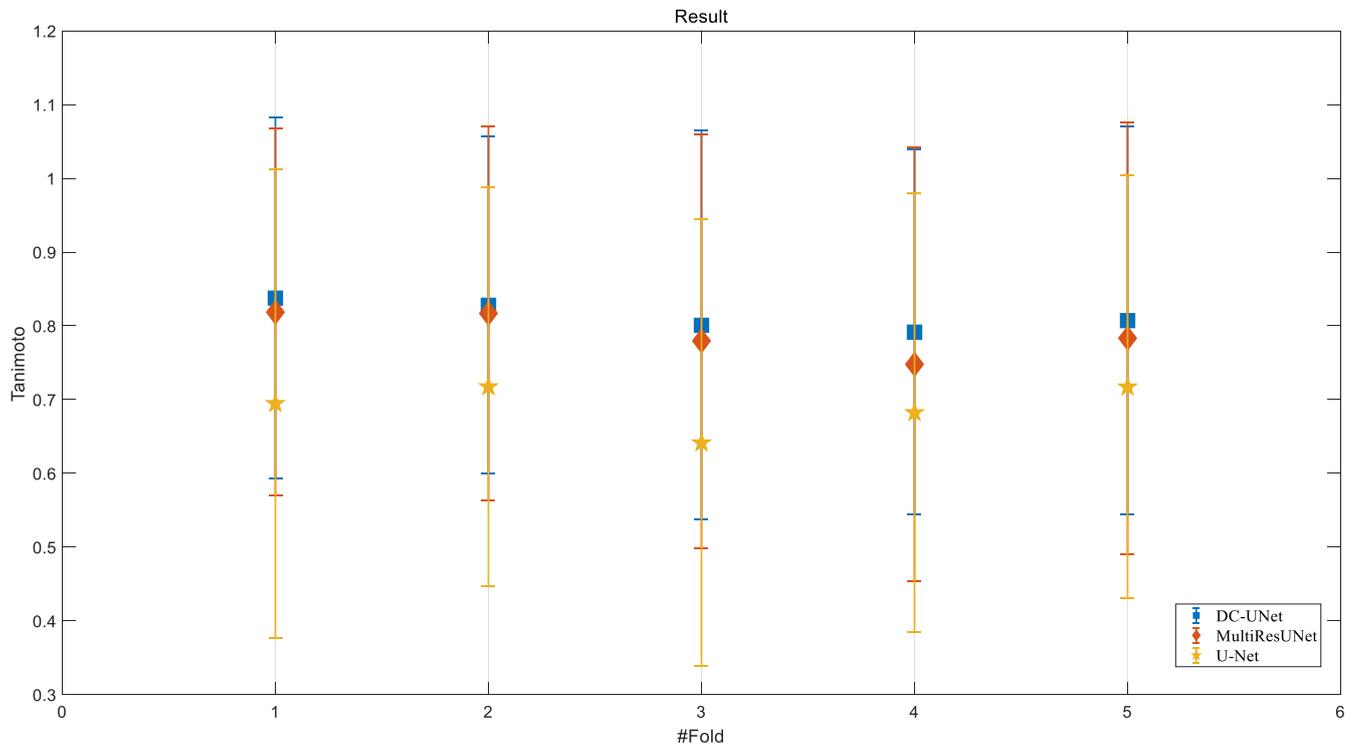
Fig. 18. Segmentation results of the models for each fold

For the endoscopy dataset, we performed 5-fold cross-validation and compared the performance of DC-UNet with MultiResUNet and the baseline U-Net. In the experiments, each model has been trained 150 epochs for each run and recorded the Tanimoto accuracy. The results of endoscopy dataset were shown in **Table 7**. The average accuracies and standard deviations for each fold are shown in **Fig. 18**.

| Model | Fold 1 | Fold 2 | Fold 3 | Fold 4 | Fold 5 | Average |
|---|---|---|---|---|---|---|
| U-Net | 74.03 | 70.81 | 67.96 | 63.26 | 71.52 | 69.52 |
| MultiResUNet | 81.82 | 80.34 | 79.57 | 74.23 | 78.66 | 78.92 |
| DC-UNet | **83.11** | **82.51** | **81.10** | **78.14** | **79.84** | **80.94** |

Table 7. The result of endoscopy through a 5-fold cross-validation. Bold values are the maximum for each case.

From the Table 8, we can find that MultiResUNet gives much better results than U-Net in this challenging dataset. The average accuracy has been improved 9.4%. However, MultiResUNet does not perform well for some challenging tasks, as shown in **Fig. 19**. The DC-UNet can successfully segment images with vague boundaries and successfully detect small objects in images, as shown in **Fig. 20**. The segmentation accuracy has been improved 11.42% to the U-Net.

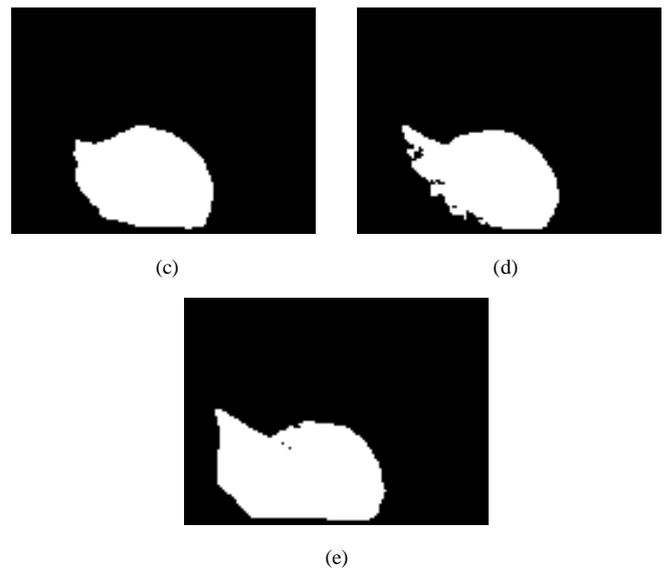

Fig. 19. Segment images with vague boundaries. (a) Original image (b) Ground truth (c) U-Net (72.25%) (d) MultiResUNet (73.04%) (e) DC-UNet (96.45%)

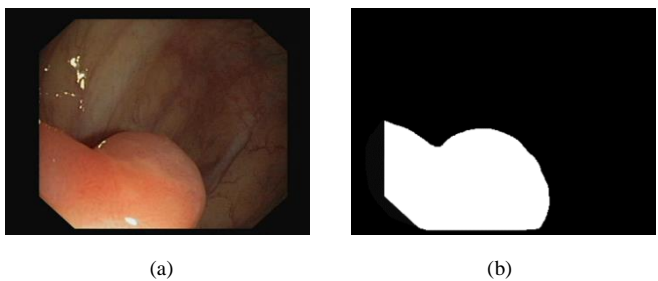

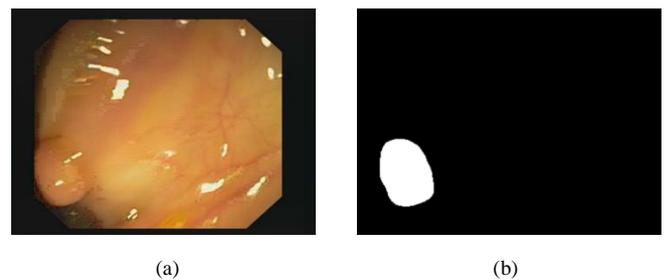

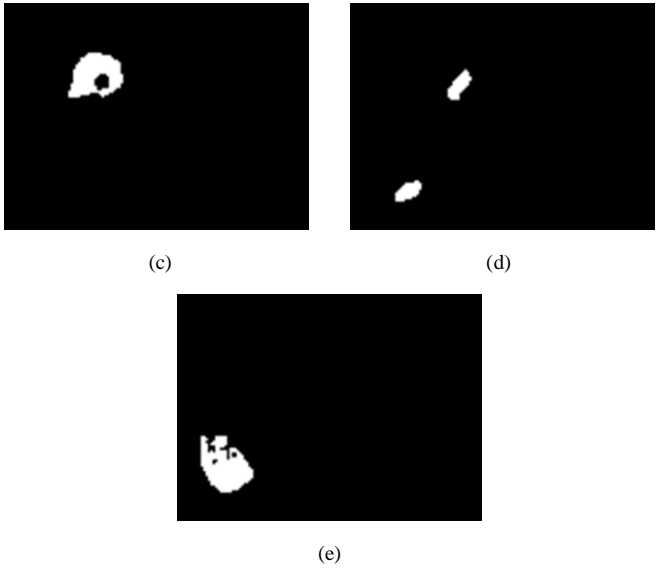

Fig. 20. Segment images with small objects. (a) Original image (b) Ground truth (c) U-Net (0%) (d) MultiResUNet (11.76%) (e) DC-UNet (69.00%)

For some easy cases in CVC-ClinicDB, results of MultiResUNet and DC-UNet are similar, but much better than classical U-Net, as shown in **Fig. 21**.

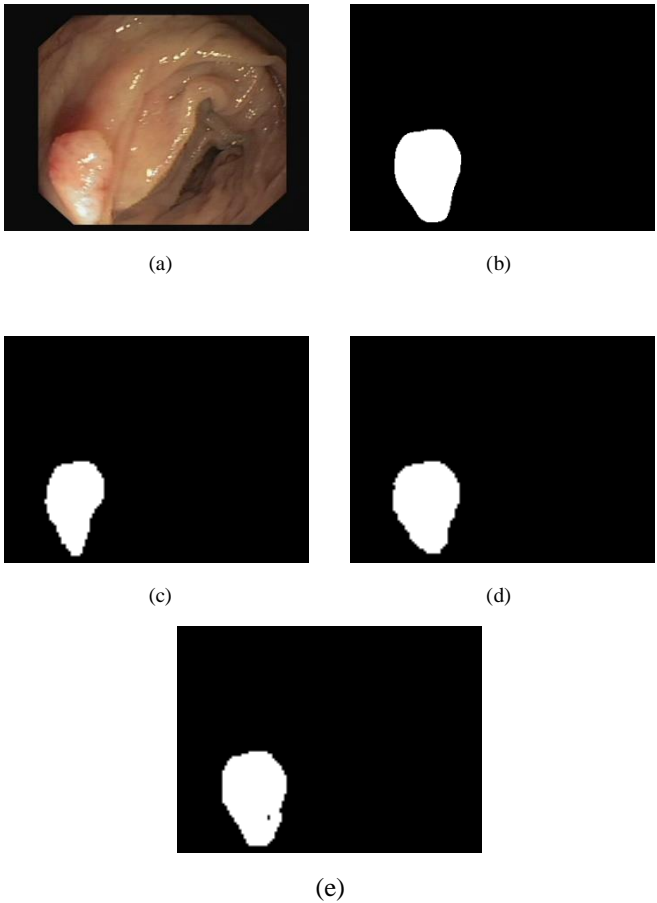

Fig. 21. Segment images with vague boundaries. (a) Original image (b) Ground truth (c) U-Net (72.07%) (d) MultiResUNet (96.44%) (e) DC-UNet (96.41%)

VI. DISCUSSION

A. *Comparison of the three models*

From these experiments, DC-UNet shows great potential in multimodal medical image segmentation. In our infrared breast dataset, DC-UNet can separate the belly and breast region even they have similar temperature and DC-UNet also provides a more accurate contour. . From the results of EM dataset as shown in **Fig. 16**, the DC-UNet shows a good robustness to noise because compared with results of MultiResUNet and U-Net, the result of DC-UNet contains less noise. In the CVC-ClinicDB dataset, DC-UNet shows a great ability, which MultiResUnet and U-Net do not have to segment small objects and vague boundaries without any data augmentation techniques.

In addition, DC-UNet is more efficient because its parameters are much less than MultiResUNet and classical U-Net. The number of parameters is related to convolutional kernel size, input's and output's channel numbers. In our DC block, each channel's filter number is half of the corresponding MultiRes block. After passing the add layer, we calculate sum of these two channels instead of concatenating. Thus, the dimension of current output layer and next input layer are half of the corresponding layer in MultiRes block. Moreover, half output dimension in DC block also leads to the filter number of Res-Paths is half of that in MultiResUNet. Based on reduced dimension of input and output, the parameters in DC-UNet are much less than MultiResUNet and U-Net. Nevertheless, it contains doubled multi-resolution features that makes the results better than compared models.

B. *Improvements and future work*

In order to get better results, data augmentation [46] like flipping, rotation and randomly cropping to enlarge the datasets and image enhancement [47] are very helpful techniques. Data augmentation operations can help models avoid overfitting during training [48]. Moreover, objects in medical images sometimes do not have clear boundaries because of poor illumination, noise and tissue properties. Thus, the histogram equalization technique, which can improve the contrast, such as CLAHE [49] would be greatly helpful.

In spite of data augmentation and image enhancement techniques, there are also potential in the dual-channel CNN architectures. In our experiments, we only use dual-channel model for segmentation. Adding more channels like blocks in ResNeXt [50] will provide more effective features, but it will cause the increment of parameters and floating points operations (FLOPs). Moreover, there are also other versions' Inception module like Inception-v4 [51] and Inception-v3 [52], which give a good idea that using asymmetric convolution to replace the original convolution kernel. For example, the $3 \times 3$ convolution operator can be replaced by a 3 × 1 convolution following a 1 × 3 convolution to minimize the parameters further.

In the future, we will test our model on more datasets. Moreover, we will study on how data augmentation and pre-processing could improve the model's performance.

VII. CONCLUSION

In this work, we analyzed the classical U-Net and the recent MultiResUNet architecture and found potential improvements. We noticed that the results of our own infrared breast dataset still have many limitations for using classical U-Net and MultiResUNet. The author of MultiResUNet paper [31] has verified that Res-Path can slightly improve the segmentation accuracy. Thus, we designed the Dual-Channel CNN block to give more effective features with less parameters to overcome those limitations. To incorporate this dual-channel CNN architecture with Res-Path, we developed a novel U-Net-like architecture--DC-UNet.

We selected two public medical datasets and our own infrared breast dataset to test and compare the performance of these three models. Each dataset contains some challenging cases. The infrared breast dataset contains small-size breast images with unclear boundaries. Some images in ISBI-2012 Electron Microscopy dataset contain many interferences like noise and other parts of cell will influence the model to recognize the boundaries. For colon endoscopy images in CVC-ClinicDB, the boundaries of polyps are very vague and hard to distinguish and the shapes, sizes, structures and positions of polyps are different. Those factors make this dataset most challenging.

For those challenging cases, the performance of DC-UNet was better than MultiResUNet and DC-UNet. Generally, for the infrared breast, ISBI-2012 and CVC-ClinicDB dataset, a relative improvement of segmentation accuracy 2.90%, 1.49% and 11.42% has been observed in using DC-UNet over U-Net. And DC-UNet also has 1.20%, 0.66% and 2.02% improvements over Multi-ResUNet. Besides higher segmentation accuracies DC-UNet achieved, the segmentation results are much closer to the ground truth by observation. As shown in **Fig. 17** and **Fig. 18**, U-Net and MultiResUNet tend to under-segment and even miss the objects completely. On the contrary, DC-UNet seems more reliable and robust. DC-UNet can detect vague boundaries and avoid the interference of noise. Even for the challenging cases, the DC-UNet shows a stronger ability to capture the fine details.

Therefore, we believed that the DC-UNet architecture can be an effective model for medical image segmentation.